\documentclass{article}
\usepackage{authblk}
\usepackage[utf8]{inputenc}
\usepackage[T1]{fontenc}
\usepackage{soul}
\usepackage{graphicx,epsfig}
\usepackage[backend=bibtex]{biblatex}
\bibliography{sample}
\usepackage{graphics,dcolumn,bm,epic, eepic,float}
\usepackage{amssymb,amsmath,amsfonts,multirow,rotate,color}
\usepackage{soul,xcolor,url}
\usepackage{hyperref}
\usepackage{tocloft}
\usepackage{morefloats}
\usepackage{geometry}
\usepackage{booktabs}

\title{Spatiotemporal variability and prediction of e-bike battery levels in bike-sharing systems}

\author[1,*,+]{Aleix Bassolas}
\author[1,2,+]{Jordi Grau-Escolano}
\author[1]{Julian Vicens}
\affil[1]{Eurecat, Centre Tecnològic de Catalunya, Barcelona, Spain}
\affil[2]{Universitat Politècnica de Catalunya, Barcelona, Spain}
\affil[*]{aleix.bassolas@eurecat.org}
\affil[+]{these authors contributed equally to this work}

\begin{document}
\maketitle

\begin{abstract}
Bike Sharing Systems (BSSs) play a crucial role in promoting sustainable urban mobility by facilitating short-range trips and connecting with other transport modes. Traditionally, most BSS fleets have consisted of mechanical bikes (m-bikes), but electric bikes (e-bikes) are being progressively introduced due to their ability to cover longer distances and appeal to a wider range of users. However, the charging requirements of e-bikes often hinder their deployment and optimal functioning.
This study examines the spatiotemporal variations in battery levels of Barcelona's BSS, revealing that bikes stationed near the city centre tend to have shorter rest periods and lower average battery levels. Additionally, to improve the management of e-bike fleets, a Markov-chain approach is developed to predict both bike availability and battery levels. This research offers a unique perspective on the dynamics of e-bike battery levels and provides a practical tool to overcome the main operational challenges in their implementation.
\end{abstract}

\section*{Introduction}

By 2022, more than 2,000 Bike Sharing Systems (BSSs) were available worldwide, with approximately 9 million bikes in circulation \cite{meddin}. These systems enhance the efficiency of public transport by improving access to stations \cite{noland2019bikesharing, kapuku2021assessing}, which in turn reduces car usage and alleviates traffic congestion \cite{ma2020bike,fan2020dockless}. Moreover, the increased use of bikes and the resulting shift in transport modes have been shown to positively impact public health by promoting physical activity \cite{otero2018health}, benefit the environment by reducing pollutant emissions \cite{clockston2021health}, and lead to economic savings \cite{ricci2015bike}.

The recent increase in data availability related to BSSs has led to greater interest in understanding the mobility patterns within these systems \cite{kou2019understanding, li2020understanding}. Studies have shown that stations with strong flows tend to be located nearby \cite{chen2017understanding}, and they often display a certain level of mesoscopic organisation within communities \cite{zaltz2013structure}. Additionally, the development of models to understand the usage of micro-mobility \cite{tran2015modeling} has improved understanding of factors such as topography \cite{kim2020anatomy}, day of the week \cite{tran2015modeling}, weather conditions \cite{elAssi2017effects, Kim2018investigation,ashqar2019modeling}, and the built environment \cite{tran2015modeling}. Furthermore, efforts have been made to develop predictive models that enhance the management of BSSs infrastructures efficiently and guide system expansion through machine learning and probabilistic approaches \cite{singhvi2015predicting, li2015traffic,chen2016dynamic, albuquerque2021machine, yang2020using, almannaa2020dynamic}. In this line, one of the major challenges in optimising BSS performance is ensuring bike availability, as stations often experience significant flow imbalances \cite{yang2016mobility, singla2015incentivizing, faghih2017empirical}. To address this, models focused on optimising rebalancing operations have been developed \cite{raviv2013static,dell2014bike,dell2018bike,chiariotti2018dynamic}.

In recent years, BSSs have undergone progressive electrification \cite{meddin, galatoulas2020spatio}, with a sharp increase in the number of e-bikes compared to their mechanical counterparts \cite{galatoulas2020spatio, bielinski2020lessons}, as e-bikes allow for longer trips and attract a broader user base \cite{simsekoglu2019role, weinert2007transition}. E-bikes have been shown to be less sensitive to factors such as longer distances, adverse weather conditions, and casualties \cite{wamburu2021ride, campbell2016factors, siman2018look}. Despite these advantages, e-bikes present additional challenges to maintaining bike availability, primarily due to low battery levels, battery malfunctions, and complications arising from high temperatures \cite{ji2014electric, florez2018development, zhang2019battery, usama2019towards}. Moreover, the scarcity of battery data at the system level complicates the development of predictive models. While several studies have implemented frameworks to predict the battery consumption of individual bikes during trips \cite{burani2022algorithm, steyn2014comparison}, most approaches at the BSS level focus on the station-level granularity \cite{zhou2023dynamic}.

This publication addresses the challenge that the state of charge poses to e-bike availability by analysing mobility and battery level data from Barcelona's BSS (Bicing) and proposing a model that predicts future bike locations with their corresponding battery levels. The structure of the paper is as follows: first, the spatial and temporal variability in average battery levels is analysed to identify regions more vulnerable to battery depletion. Next, a Markov chain approach is implemented to predict the future locations of bikes. Subsequently, the parameters in the equations describing battery power consumption are optimised. Finally, by combining the models presented in the two previous sections, a framework is provided to support the management of e-bike fleets by projecting future scenarios.

\section*{Results}

\subsection*{Assessing e-bikes' battery levels and usage patterns}

This research analyses the bike usage and battery levels of Barcelona's BSS \cite{soriguera2018simulation, winslow2019bicycle, grau2024cycling}, using data supplied by its operator, Bicing, which includes trip and battery information:

\begin{itemize}
\item The mobility data spans between April 2019 and December 2022, comprising 53 million trips, of which 62\% were performed with m-bikes and 38\% with e-bikes. Each trip record includes the station of origin and destination, the start and end date of the trip, a bike identifier, the bike model (m-bike or e-bike), and an anonymised user identifier \cite{grau2024cycling}. In this work, we focus specifically on the mobility of e-bikes.

\item To analyse e-bike battery consumption, two datasets were used: one with snapshots taken every 8 hours over the entire year of 2022, and another with snapshots taken every 30 minutes over three weeks from 29 March to 18 April 2023 (Figures S1 and 3
). The relevant fields in both datasets include percentage, voltage, current, and temperature of the battery, bike station, timestamp, bike status (operative, low battery, or inoperative for several reasons such as maintenance), and location (at station, in use, or at the workshop). A detailed analysis of these datasets is provided in Sections S1 and S2.
\end{itemize}

Additionally, trip data from March and April 2023 was available to assess battery consumption in the 30-minute dataset.

\begin{figure}[!htbp]
    \centering
    \includegraphics[width=0.9\textwidth]{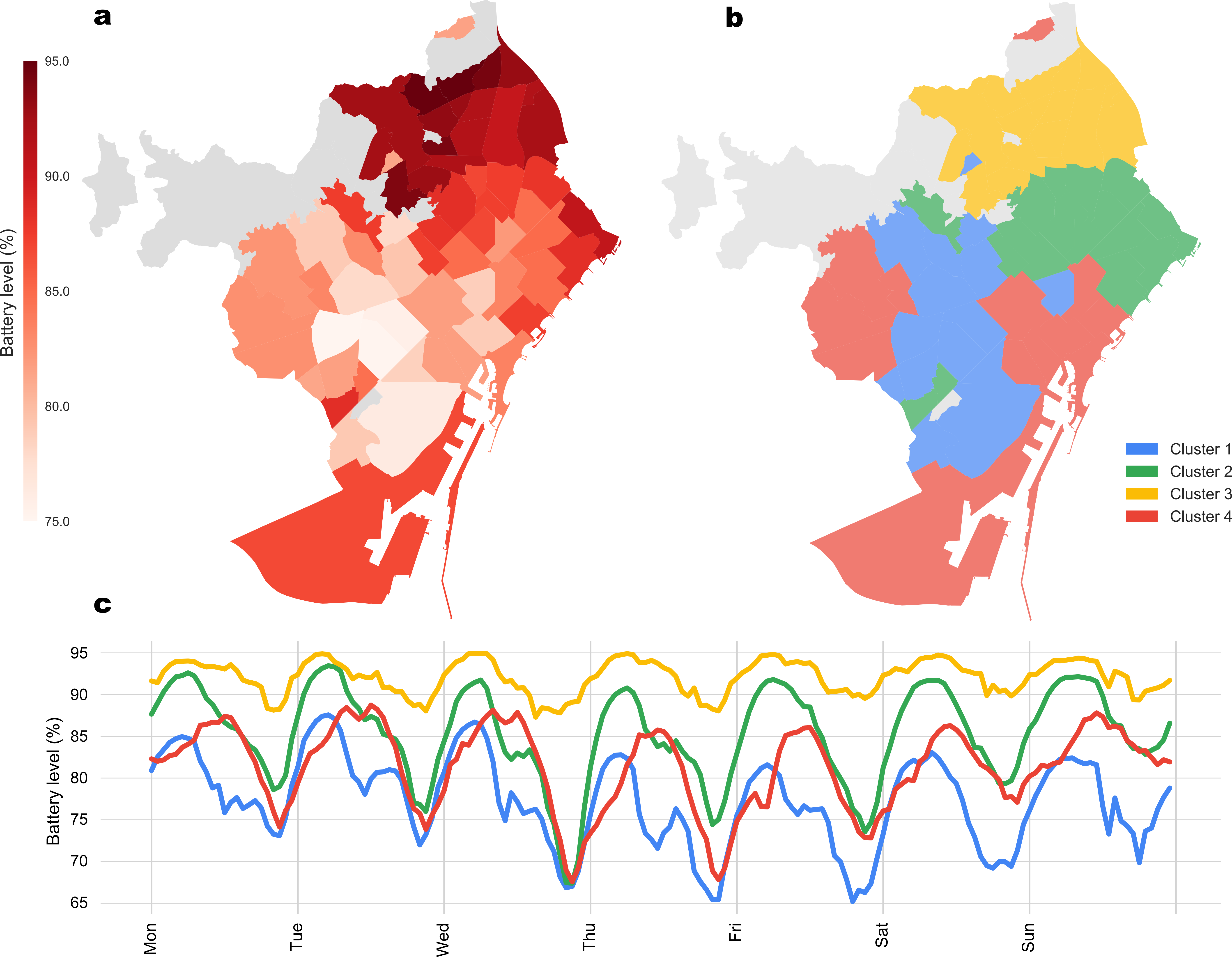}
    \caption{\textbf{Analysis of average battery levels and temporal clustering}. \textbf{a.} Average battery levels across neighbourhoods. The average for each neighbourhood is computed using the battery values of all the bikes in that neighbourhood. \textbf{b.} Clustering of the neighbourhoods according to the temporal pattern of battery levels using K-means into four distinct clusters. \textbf{c.} Temporal series of average battery levels for each cluster. The neighbourhoods in grey do not have bike-sharing stations.}
    \label{battery_overiew}
\end{figure}

The average battery levels per neighbourhood show significant variability across neighbourhoods (Figure~\ref{battery_overiew}a, Figure S3). To facilitate the analysis of all the temporal series of average battery levels derived from each neighbourhood, K-means clustering was performed, resulting in four distinct clusters (Figure~\ref{battery_overiew}b, Figure S4). The central and southern neighbourhoods, which fall into Clusters 1 and 4, have the lowest battery levels, exhibiting steep declines in battery levels during business hours with recovery during nighttime (Figure~\ref{battery_overiew}c). This is likely due to the high levels of industrial and commercial activity, as well as the dense populations in these areas. Additionally, it is worth noting that some neighbourhoods in Cluster 4 have slower charging speeds at night, likely due to a stronger nightlife presence. In contrast, northern neighbourhoods in Clusters 2 and 3 are mainly residential with less commercial activity, resulting in higher and more stable battery levels throughout the day.

\begin{figure}[!htbp]
    \centering
    \includegraphics[width=0.9\textwidth]{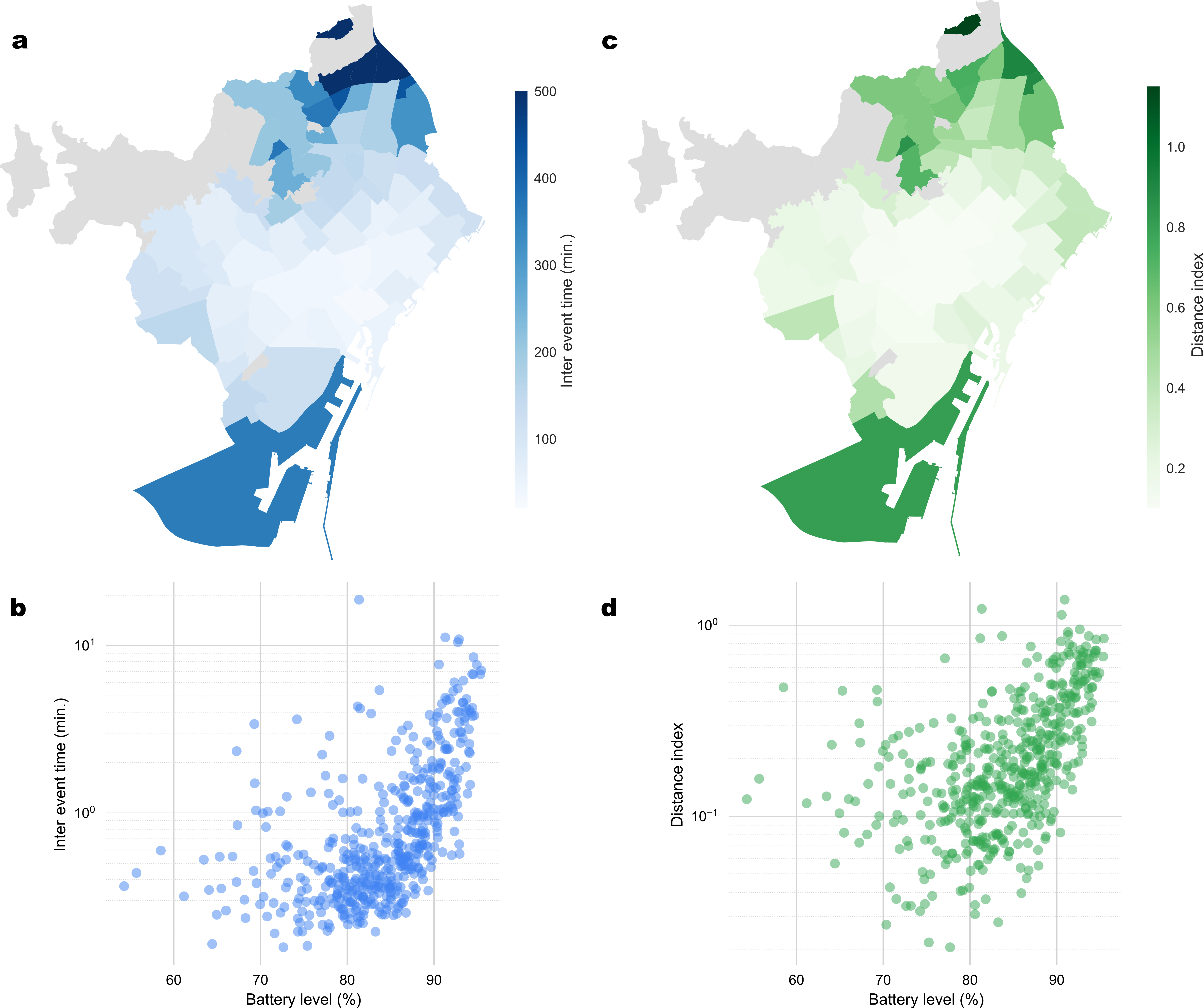}
    \caption{\textbf{Analysis of inter-event times and distance index for e-bikes}. \textbf{a.} Average inter-event time between trips across neighbourhoods. The average for each neighbourhood is computed using the inter-event time values of all bikes, with outliers removed using the IQR method at the station level. \textbf{b.} Relation between average battery levels and inter-event time at the station level, with a Pearson correlation of 0.39 (p-value<0.001). \textbf{c.} Distance index per neighbourhood, calculated by first computing the index at the station level to accurately reflect trip distances between stations, and then averaging it at the neighbourhood level. \textbf{d.} Relation between average battery levels and the distance index at the station level, with a Pearson correlation of 0.47 (p-value<0.001). The neighbourhoods in grey do not have bike-sharing stations.}
    \label{correlations}
\end{figure}

One factor directly related to average battery levels is the average time between trips for each bike, which can also be referred to as inter-event time. With longer inter-event times, e-bikes have more opportunity to recharge, resulting in higher battery levels. In Figure~\ref{correlations}a, central and southern neighbourhoods (Clusters 1 and 4) have not only lower average battery levels but also shorter inter-event times, indicating a higher turnover of e-bikes due to frequent trips. This high turnover aligns with the higher density of users and increased mobility demand in commercial and densely populated areas. However, Zona Franca, located in the southernmost area of the city, presents a distinct case due to its industrial nature and significantly low population density (1 inhabitant per hectare \cite{bcn_population_density}), which results in low e-bike usage. Conversely, northern neighbourhoods (Clusters 2 and 3) exhibit longer inter-event times, reflecting less frequent e-bike usage and consequently higher and more stable battery levels despite some of them being in high altitudes. Additionally, a Pearson correlation analysis was conducted between average battery levels and inter-event time at the station level (Figure~\ref{correlations}b), revealing a correlation value of 0.39. This moderate positive correlation suggests that stations with longer inter-event times tend to have higher average battery levels, further supporting the neighbourhood-level results.

Apart from the inter-event time, two additional factors can affect the average battery levels at a station: the origins of the trips that the station receives and the number of trips coming from each origin. Trips originating from farther stations tend to deplete the battery more. Therefore, stations that receive a greater number of trips from distant origins will typically have lower average battery levels compared to those receiving trips from nearby stations. To quantify the impact of trip distances on battery levels, a distance index (\(DI_j\)) was calculated for each station \( j \). This index accounts for both the distance of each trip and the frequency of trips from each origin station. Specifically, the distance of each trip (\( d_{i,j} \)) from station \( i \) to station \( j \) is weighted by the inverse of the number of trips from station \( i \) (\( n_i \)). These weighted distances are then summed for each destination station \( j \) and normalised by the total number of trips received by station \( j \) (\( \sum_{i} T_{ij} \)) to account for the overall activity level at the station:

\begin{equation}
  DI_j = \frac{\sum_{i} \frac{d_{i,j}}{n_i}}{\sum_{i} T_{ij}}
\end{equation}

The distance index was computed for all stations, and averaged over each neighbourhood (Figure~\ref{correlations}c). As shown, neighbourhoods in the northern areas tend to have a higher distance index, indicating that trips to these neighbourhoods often originate from farther stations and that there are fewer incoming trips overall. This combination results in higher average battery levels. Conversely, neighbourhoods in the central and southern parts of the city exhibit a lower distance index, suggesting that trips to these areas typically come from nearby stations and that there are more incoming trips, leading to lower battery levels. Zona Franca, as previously mentioned, deviates from this trend due to its unique industrial nature and low population density. The Pearson correlation between the distance index and average battery levels at the station level revealed a value of 0.47 (Figure~\ref{correlations}d), indicating a moderate positive correlation. This suggests that stations farther from others and receiving a lower number of trips tend to have higher average battery levels. Additionally, no significant collinearities were found between the inter-event time and the distance index (Figure S5 and Table S1).

\subsection*{Predicting e-bike locations with a Markov chain-based approach}

To predict bike locations, a probabilistic model based on a Markov chain approach has been developed \cite{lu2013approaching,huang2015predicting}. In this model, each bicycle of the system is modelled independently, with transitions based on the probability of movement from its current station. Given a starting location for a bike and a forecasting horizon $\Delta t$, the algorithm follows the sequence below:

\begin{itemize}
    \item An inter-event time is sampled from the distribution $I_i^{w,h}$ to simulate the resting time of a bike given a station $i$ on a specific day type $w$ and hour $h$. The days of the week are grouped into four categories: Working days (Monday to Thursday), Friday, Saturday, and Sunday. The distribution $I_i^{w,h}$ is calculated by determining the time elapsed between a bike’s arrival at station $i$ and its next departure.
    
    \item If the inter-event time exceeds the desired $\Delta t$, the process stops, and the final location of the bike will correspond to its current station. 
    
    \item If the forecasting horizon is greater than the inter-event time, a destination $j$ is sampled based on the probability $M_{ij}^{w,h}$, given a weekday $w$, hour $h$, and origin station $i$. This destination then becomes the new location of the bike, and the average travel time between the station pairs, $\tau_{ij}$, is elapsed. The destination probability is calculated using the number of trips $T_{ij}^{w,h}$ between $i$ and $j$ from the 2022 trip data, as follows:
    
        \begin{equation}
            M_{ij}^{w,h}=\frac{T_{ij}^{w,h}}{\sum_{\forall j} T_{ij}^{w,h}},
        \end{equation}

    \item If the total time elapsed after completing the trip exceeds the desired $\Delta t$, the final predicted location will be station $j$. Otherwise, the process starts again.
    
\end{itemize}

For each bike, 200 realizations are performed to capture a representation of all possible scenarios. The main result is the probability of ending up at station $j$ after a time period $\Delta t$, denoted as $f^{w,h}_{i,j}(\Delta t)$. This probability is calculated as the number of times a bike ends up at station $j$, divided by the total number of iterations. To aggregate the results at the system level and determine the number of bikes available per station, the probabilities $f^{w,h}{i,j}(\Delta t)$ are summed for each station $j$, which may result in non-integer values for the number of bikes at each destination.

\subsection*{Modelling e-bike battery levels}

To model the battery levels of bikes, a linear increasing term was applied when docked, and a linear decreasing term during trips. The main variables available to compute battery consumption during a trip are its duration and the distance between the origin and destination stations. Battery consumption is calculated by multiplying the trip duration by the instantaneous power consumption. To find the optimal values for the coefficients in the power consumption equation for individual e-bikes \cite{burani2022algorithm,steyn2014comparison}, as well as the charging speed, the Optuna Framework was employed \cite{akiba2019optuna} (see Methods). The 30-minute dataset was split into 75\% for training and 25\% for validation.

\begin{figure}[!htbp]
  \begin{center}
  \includegraphics[width=0.9\textwidth]{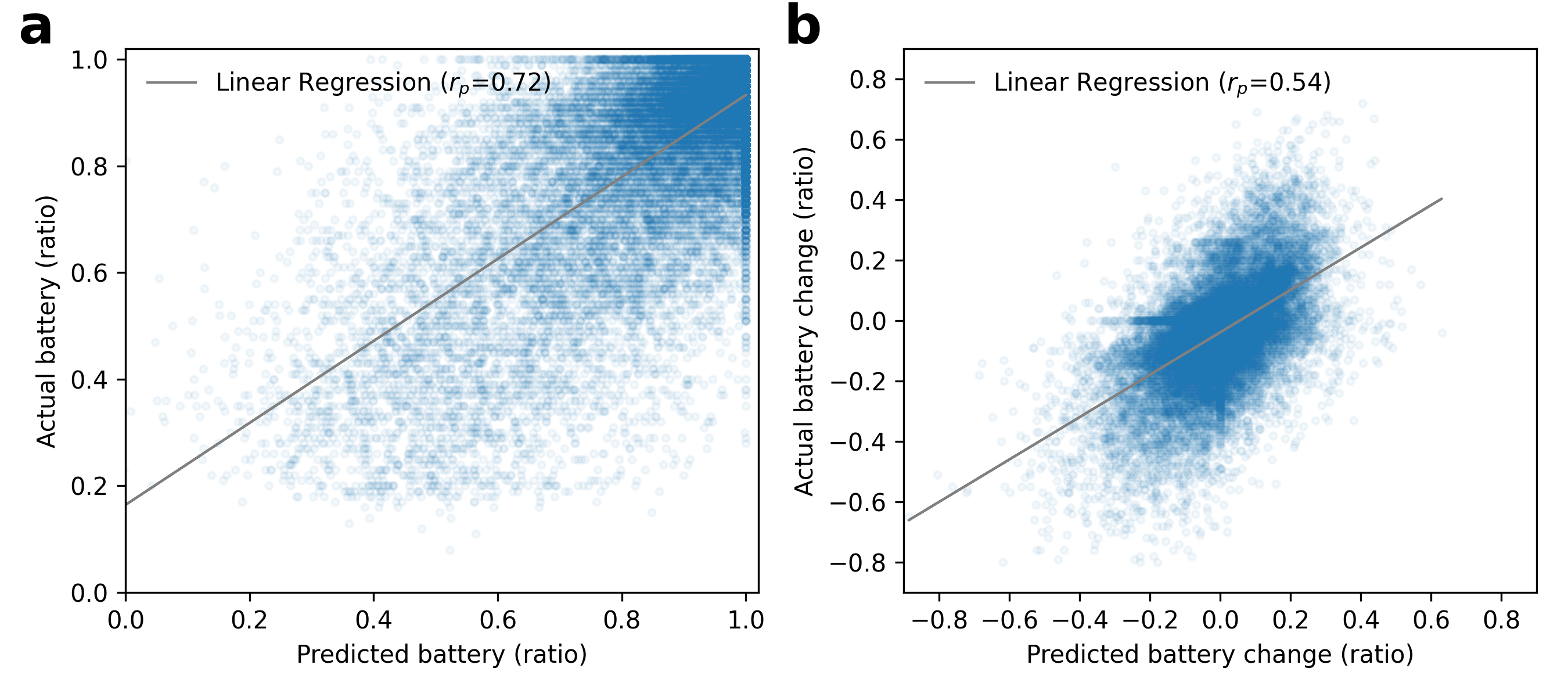}
  \end{center}
  \caption{\textbf{Battery model predictions for the validation set of the 30-minute dataset.} \textbf{a.} Observed battery level as a function of the predicted values. \textbf{b.} Battery variation observed as a function of the predicted values. Each point corresponds to one battery-level observation. The Root Mean Squared Error (RMSE) is $0.0178$ and the Pearson correlation coefficient is significant (p-value<0.001) in both cases. The linear fit is shown in grey.} \label{fig:battery_fit}
\end{figure}

The fit between the observed and expected battery levels, using the optimal parameters, is shown in Figure \ref{fig:battery_fit}. In Figure \ref{fig:battery_fit}a, the predicted battery percentage is displayed, while in Figure \ref{fig:battery_fit}b, the predicted battery variation is presented. Despite the dispersion, a significant correlation is observed, with values closely aligned to the diagonal. Factors such as user heterogeneity, missing route information, or bikes not charging while docked could contribute to the dispersion. The parameters were also validated using the 8-hour dataset, with similar correlation values, as shown in Figure S6 (Section S3).

This battery model can have two key applications. The first enables the assessment of battery consumption and destination probabilities for individual trips. The second integrates the Markov chain-based approach with the battery model to predict both e-bike locations and battery levels across the entire fleet.

\subsubsection*{Single trip application}

\begin{figure}[!htbp]
  \begin{center}
  \includegraphics[width=0.9\textwidth]{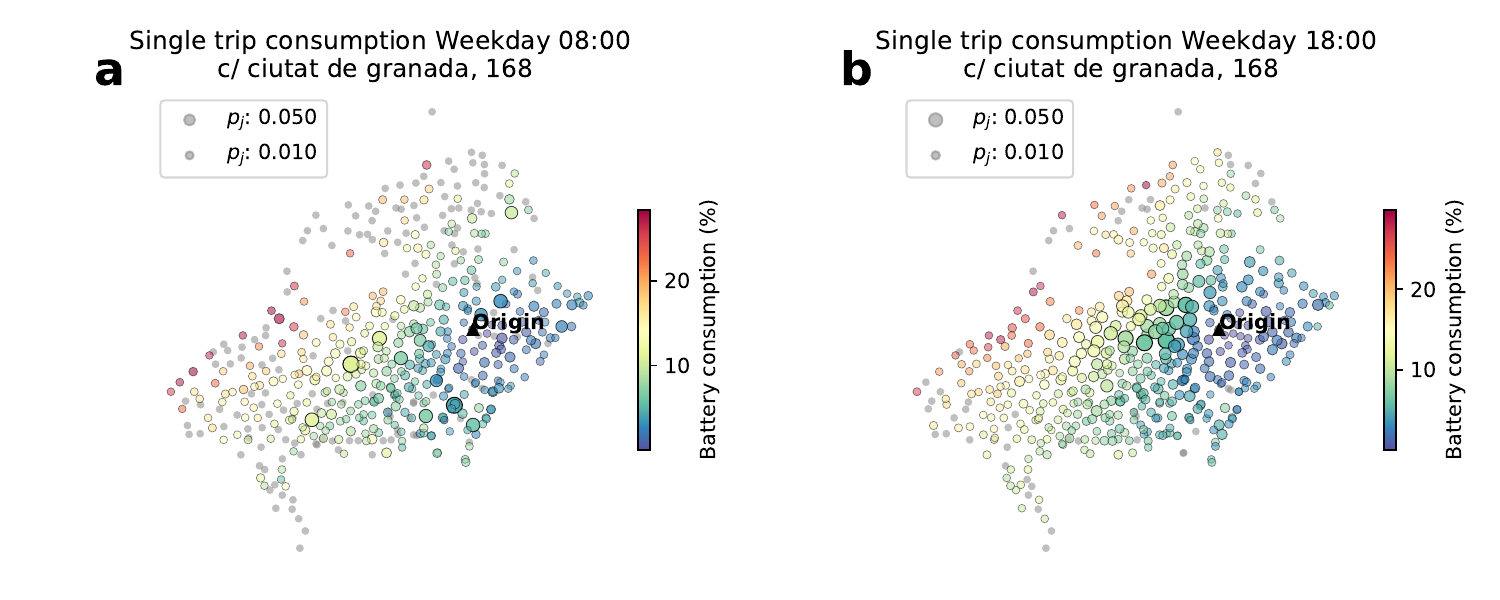}
  \end{center}
  \caption[Single trip location and battery prediction]{\textbf{Single trip location and battery prediction}. Probability of finding a bike at a station and battery consumption of the trip when departing from the \textit{St. Ciutat de Granada, 168} station on a weekday at \textbf{a.} 08:00 and \textbf{b.} 18:00. The origin station is highlighted with a triangle. The size of the dots corresponds to the probability of having that station as the destination and color corresponds to the battery consumed by the bike. The grey points correspond to destination stations with no trips. } \label{fig:single_trip}
\end{figure}

Evaluating short-term scenarios for a single bike can be achieved by coupling the battery model with the distribution of possible destinations $P_{ij}^{w,h}$. In Figure \ref{fig:single_trip}, given \textit{St. Ciutat de Granada, 168} as the origin station, the probability of arriving at each destination is represented by the size of the dots, while the battery consumption for each bike trip is indicated by the colour. The probability of destinations changes significantly between 8:00 and 18:00 despite the same origin station. At 18:00, trips tend to be shorter and head towards nearby stations, while morning trips are longer and converge towards a few work-related hot spots. Additionally, battery consumption varies notably based on the elevation changes along the route, with trips going downhill consuming less battery over the same distance. The altitude of each station is provided in Figure S7 \cite{opentopodata}.

\begin{figure}[!htbp]
  \begin{center}
  \includegraphics[width=0.9\textwidth]{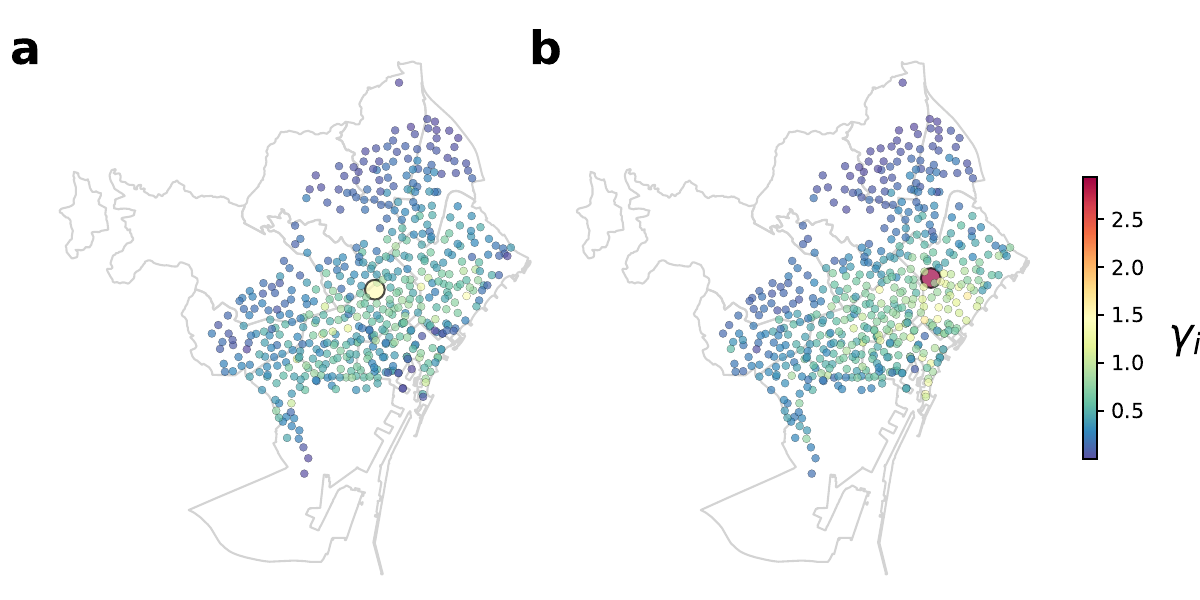}
  \end{center}
  \caption{\textbf{Impact index of the stations at different hours.} Impact index of the stations at \textbf{a.} 08:00 and \textbf{b.} 18:00. The station with higher impact at each hour is highlighted with a larger dot size. In the morning, the \textit{St. Provença, 125} station has the highest impact, while in the afternoon, \textit{St. Ciutat de Granada, 168} has the greatest impact.} \label{fig:impact}
\end{figure}

Furthermore, an impact index can be computed to assess the influence of each origin station on battery levels. This index depends on the probability of each destination given an origin station, the battery consumption of each possible trip, and the average hourly trips departing from station $i$ to other stations. The impact index can be written as:

\begin{equation}
\gamma^{w,h}_i=\langle T^{w,h}_i \rangle P^{w,h}_{ij} \Delta B_{ij}
\end{equation} 

where $\langle T^{w,h}_i \rangle$ represents the average number of trips departing from station $i$, $P^{w,h}_{ij}$ is the probability that a trip generated in station $i$ ends in station $j$, and $\Delta B_{ij}$ is the battery consumed by a trip from $i$ to $j$. An example of the impact index $\gamma$ at 8:00 and 18:00 on a regular weekday is shown in Figure \ref{fig:impact}. Due to the variation in trip patterns captured by $P^{w,h}_{ij}$, the stations with the highest impact change throughout the day.

\subsection*{Coupling the Markov chain and the battery models}

\begin{figure}[!htbp]
  \begin{center}
  \includegraphics[width=0.9\textwidth]{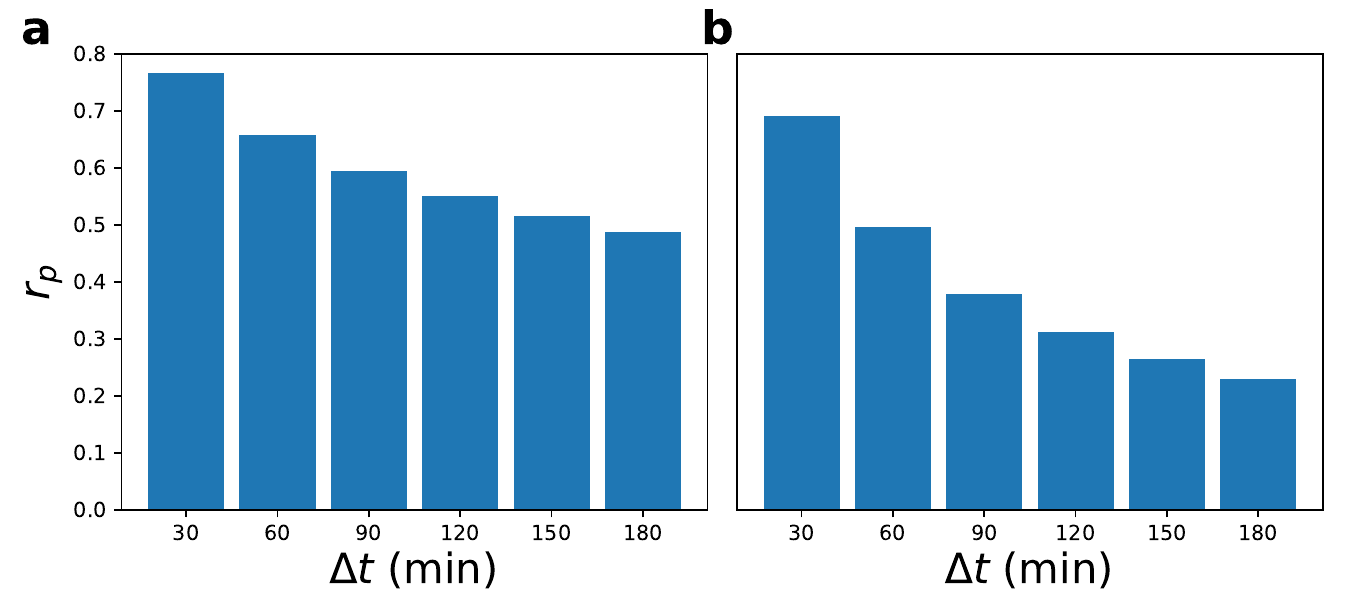}
  \end{center}
  \caption{\textbf{Correlations for the Markov chain model over various time horizons.} \textbf{a.} Pearson correlation coefficient between actual and predicted bike availability per station as a function of $\Delta t$. \textbf{b.} Pearson correlation coefficient between actual and predicted average battery level per station as a function of $\Delta t$.} \label{fig:markov_correlations}
\end{figure}

The adjusted battery consumption function can be integrated into the Markov chain approach to provide joint predictions of both location and battery levels over multiple trips, either at the single-bike or BSS level. Inter-event times are used to calculate the charging percentage, while the distance and duration of trips are used to calculate energy consumption. To validate this approach, systematic simulations were performed using the initial conditions from the 2023 mobility data and 30-minute battery data (Figure \ref{fig:markov_correlations}). This included the average number of available bikes and average battery levels per station. Simulations were performed across various initial scenarios and time horizons to demonstrate the capabilities of the coupled model. Additionally, to prevent data leakage, the probabilities of the Markov model were based on data from the year 2022.

\begin{figure}[!htbp]
  \begin{center}
  \includegraphics[width=0.9\textwidth]{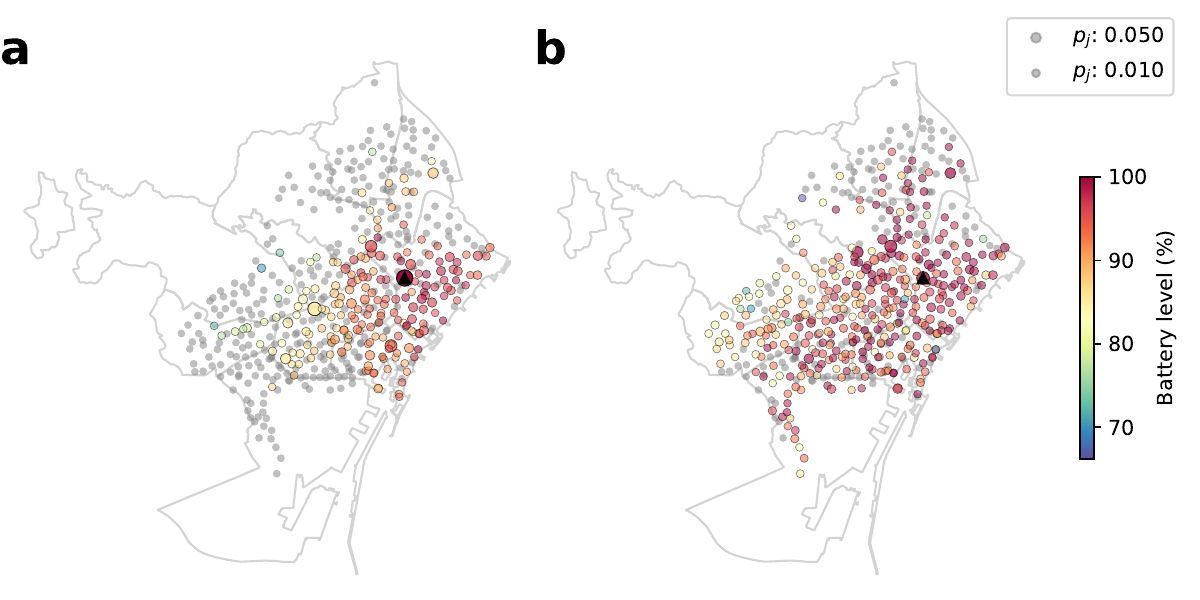}
  \end{center}
  \caption{\textbf{Bike location and average battery level predicted with the Markov model.} Probability of finding a bike and average battery levels after \textbf{a.} 30 and \textbf{b.} 120 minutes for trips starting from the Glòries station (St. Ciutat de Granada, 168) starting at 08:00. The origin station is highlighted with a triangle.} \label{fig:markov_trips}
\end{figure}

First, it is possible to compute the probability of finding a bike at a given destination and its battery level as a function of its origin, following one or more trips within a time period $\Delta t$. The final battery levels are determined by averaging over all the Markov iterations. As an example, Figure \ref{fig:markov_trips} shows the probability and battery levels of bikes departing from the station at Glòries (\textit{St. Ciutat de Granada, 168}) at 08:00 after \textbf{a.} 30 and \textbf{b.} 120 minutes. The role of distance from the origin is evident in the short-term predictions, with lower battery levels at greater distances and inclinations. However, this effect diminishes over longer time windows, as multiple trips can occur before reaching a station. For longer periods, a greater variety of final destinations is also possible.

\begin{figure}[!htbp]
  \begin{center}
  \includegraphics[width=0.9\textwidth]{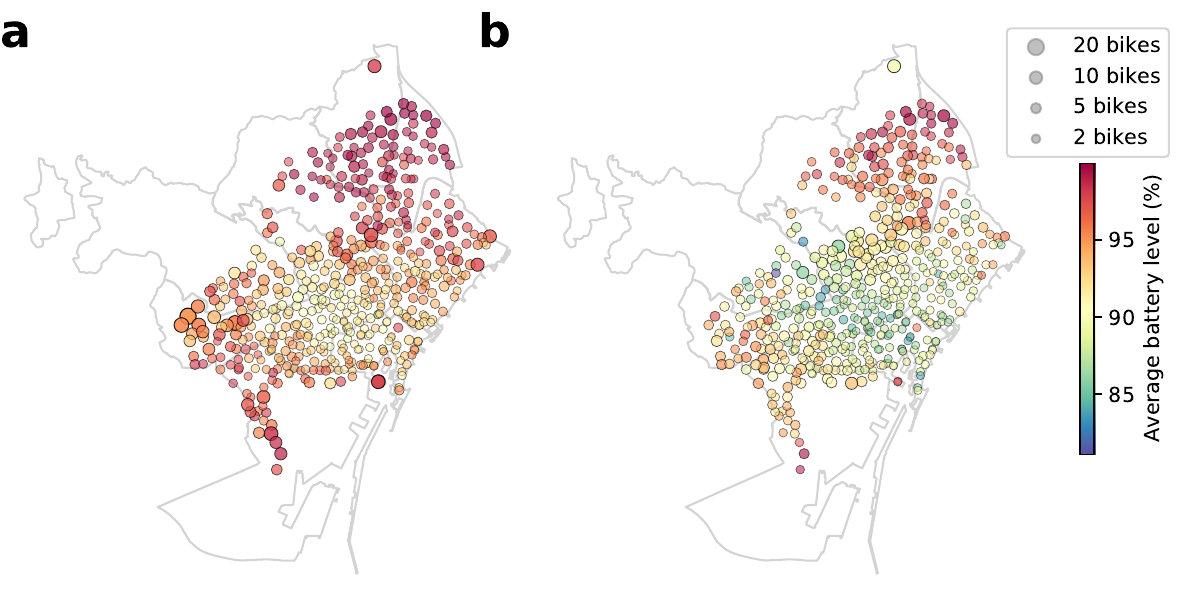}
  \end{center}
  \caption{\textbf{Bike availability and battery levels predictions at the BSS level after 120 minutes, using the Markov chain approach combined with the battery equations.} Prediction of available bikes and their average battery levels after 120 minutes of simulation, starting at \textbf{a.} 08:00 and \textbf{b.} 18:00.} \label{fig:markov_battery}
\end{figure}

As a second application, the Markov chain approach allows to transition from bike-level to system-level predictions. To calculate the number of available bikes per station across the entire fleet, the probabilities $f^{w,h}_{i,j}(\Delta t)$ for each destination station $j$ are summed, which may result in non-integer values. Battery levels, on the other hand, are calculated by multiplying each bike’s probability by its corresponding battery level and then dividing by the number of available bikes. Figure \ref{fig:markov_battery} displays the number of available bikes and the average battery level after two hours of simulation for two different starting times, 08:00 and 18:00. Stations near the city centre exhibit lower battery levels, consistent with the data analysis results. Meanwhile, the northern part of the city, which experiences lower usage rates, shows higher battery levels. Additionally, lower battery levels are observed in the afternoon due to increased battery consumption throughout the day.

\section*{Limitations}

The main constraints of this study are the limited amount of data and the lack of detailed trip information. The first dataset has an 8-hour resolution, which complicates accurate model training and evaluation. Although the second dataset offers a 30-minute resolution, it is limited to two weeks and does not include bikes in transit.
In addition to the battery data, the incomplete trip data further complicates model development. Information on the exact routes taken by users is unavailable, making it difficult to estimate battery consumption for round trips. Additionally, details such as the bike's condition (e.g., tyre inflation) or relevant rider characteristics (e.g., weight or height) are also missing.

\section*{Discussion}

The analysis revealed spatiotemporal variations in e-bike battery levels, with lower levels observed in the evenings due to daily usage, compared to higher levels in the mornings. Northern neighbourhoods tended to show higher battery levels, likely due to lower usage and longer resting times. The inter-event times and distance to other stations significantly correlate with the station's battery levels.

The presented methodology combines a traditional probabilistic framework with heterogeneous data sources, resulting in a multifaceted tool capable of improving the real-time management of micro-mobility services. Although some challenges remain due to the lack of on-route data, such as speed or location, the battery function has been validated. This function, when combined with a Markov chain approach, provides a validated prediction for the availability of e-bikes, either due to an imbalance between stations, or to an overused battery. Further improvements could be made, particularly if more detailed data were available. Access to GPS information on the routes taken would likely improve power consumption estimates, while higher-resolution data could enhance battery predictions through machine learning approaches.

With the progressive electrification of bike fleets worldwide, predictions of availability alone may be insufficient, as docked bikes with low batteries remain unusable. The presented approach opens the door to developing microscopic models that provide joint predictions of bike availability and battery levels. This framework could support BSS providers in optimising bike availability and identifying stations at higher risk of disrupting urban mobility.

\section*{Methods}

\subsection*{Mathematical formulation of the battery consumption}

The battery levels of the Bicing e-bikes are modelled with a method based on the fundamental laws of motion. The instantaneous power consumed while riding is derived from the need to overcome three main forces opposing the motion: gravity resistance ($F_g$), rolling resistance ($F_r$), and aerodynamic drag ($F_a$) \cite{burani2022algorithm, steyn2014comparison}.

The gravity resistance $F_g$ is given by:
\begin{equation}
F_g=g \cdot m \cdot \sin(arctan(s)),
\end{equation}
where $g$ is the gravitational acceleration, $m$ is the combined mass of the rider and bike, and $s$ represents the slope. Similarly to \cite{burani2022algorithm}, $F_g=0$ when $F_g\leq 0$.

Rolling resistance $F_r$ is given by:
\begin{equation}
F_r={C_{rr}} \cdot g \cdot m  \cdot \cos(arctan(s)),
\end{equation}
where $C_{rr}$ is the rolling resistance coefficient.

Finally, the aerodynamic drag $F_a$ is given by:
\begin{equation}
F_a=0.5 \cdot C_d \cdot A \cdot \rho \cdot (v+w)^{2},
\end{equation}
where $C_d$ denotes the drag coefficient, $A$ is the frontal area, $\rho$ is the air density, and $v$ and $w$ are the bike and wind speeds, respectively. 

The instantaneous power consumed is then given by:
\begin{equation}
P(t)=vF,
\end{equation}
Assuming no energy losses during the process, the total power consumed over a trajectory of duration $t$ is given by $tv(F_g+F_r+F_a)$. As the instantaneous speed information is unavailable, the speed is approximated by dividing the total distance travelled by the trip time. Consequently, the energy consumed in a trip is given by $d_{ij}(F_g+F_r+F_a)$, where the distance travelled is approximated using the OpenRouteService API \cite{openrouteservice}, which recommends bicycle routes based on OpenStreetMap data \cite{OpenStreetMap}. To convert power consumed into a battery percentage, and taking into account that the e-bikes only provide support to the rider when pedalling, a constant $K_{\rm trip}$ has been added by multiplying the whole equation $K_{\rm trip}d_{ij}(F_g+F_r+F_a)$.

On the other hand, the rate of battery recharge during rest periods is modelled as:
\begin{equation}
\Delta B_{\rm rest}=\frac{K_{\rm rest}t_{\rm rest}}{60}, \label{charge}
\end{equation}

where $K_{\rm rest}$ represents the charging constant, and $t_{\rm rest}$ is the rest time. The inverse of $K_{\rm rest}$ corresponds to the time required to fully charge a bike from $0\%$ to $100\%$. The division by $60$ is included to express $K_{\rm rest}$ in units of $min^{-1}$. All variables are represented in the International System of Units.

\subsection*{Optimisation of the battery consumption function}

The battery consumption during a trip can be approximated by an expression that groups the relevant constants. This expression is written as:

\begin{equation}
\Delta B_{\rm trip}=K_{\rm trip} d_{ij}(K_a sin(arctan(s))+ K_b cos(arctan(s)) + K_c v^2), \label{consumption}
\end{equation}

where

\begin{align*}
& K_a=g \cdot m, \\
& K_b={C_{rr}} \cdot g \cdot m, \\
& K_c=0.5 \cdot C_d \cdot A \cdot \rho.
\end{align*}

Although the constant $K_{\rm trip}$ is not strictly required mathematically, it allows us to explore feasible values for $K_a$, $K_b$, $K_c$. Apart from $g$, the constants involved in the computation of $K_a$, $K_b$, and $K_c$ are unknown and may vary depending on factors such as rider characteristics, weather conditions, bike status, and terrain type. Therefore, an optimisation process is necessary to determine the optimal values of $K_a$, $K_b$, and $K_c$ that enable accurate battery consumption predictions, even when the constants of every trip are not known precisely.

To perform the optimisation, a different range of values is considered for each constant. The mass $m$ is assumed to range between 79 kg and 119 kg, considering the bike's weight of 29 kg. Based on previous studies \cite{steyn2014comparison, bertucci2013evaluation}, the rolling resistance coefficient can range between $0.001$ and $0.003$, the drag coefficient between $0.6$ and $0.8$, and the effective area between $0.4$ and $0.6$. The air density $\rho$ is fixed at $1.225kg/m^3$, corresponding to a temperature of $15^{\circ}C$ at sea level. Given these constants' value ranges, the following limits are defined for the hyper-parameter optimisation: $K_a = [774.2, 1166.2]$, $K_b = [0.7742, 3.4986]$, and $K_c = [0.147, 0.294]$. The range for $K_{\rm rest}$ is set between $[0.002083, 0.055]$, corresponding to charging times from 480 to 180 minutes, respectively. The range for $K_{\rm trip}$ is set between $[1 \cdot 10^{-8},1]$.

The Optuna framework was utilised for hyper-parameter optimisation, using 75\% of the three-week 30-minute dataset for training and 25\% for validation. The optimal values obtained were:

\begin{align*}
& K_a = 911.139 \\
& K_b = 3.497 \\
& K_c = 0.292 \\
& K_d =1.644 \cdot 10^{-6} \\
& K_{\rm rest} = 480
\end{align*}

\section*{Funding}
This work has been funded by CIDAI (Center for Innovation in Data Tech and Artificial Intelligence) through the Catalan Government's CATALONIA.AI programme. J.G.E is a fellow of Eurecat’s “Vicente López” PhD grant program.

\section*{Acknowledgements}

We thank Daniel Santanach, coordinator of CATALONIA.AI programme. We thank Marco Orellana from CIDAI (Centre of Innovation for Data Tech and Artificial Intelligence). We want to thank Faustino Corchero from Barcelona de Serveis Municipals and Roger Junqueras, Irene Giménez and all collaborators from Pedalem-Bicing. 

\section*{Author contributions statement}

A.B., J.G.E and J.V. conceptualised the study. A.B. developed the methodology. A.B. and J.G.E. performed data analyses and models. A.B. and J.G.E. wrote the original draft; and all authors critically discussed the results, revised the paper and approved the final manuscript.

\printbibliography

@misc{meddin,
    author= {{Meddin}},
    title = {The Meddin Bike-sharing World Map},
    url =  {https://bikesharingworldmap.com},
    note = {Accessed: 2024-06-30}
}

@article{noland2019bikesharing,
  title={Bikesharing trip patterns in New York City: Associations with land use, subways, and bicycle lanes},
  author={Noland, Robert B and Smart, Michael J and Guo, Ziye},
  journal={International journal of sustainable transportation},
  volume={13},
  number={9},
  pages={664--674},
  year={2019},
  publisher={Taylor \& Francis}
}

@article{kapuku2021assessing,
  title={Assessing and predicting mobility improvement of integrating bike-sharing into multimodal public transport systems},
  author={Kapuku, Christian and Kho, Seung-Young and Kim, Dong-Kyu and Cho, Shin-Hyung},
  journal={Transportation research record},
  volume={2675},
  number={11},
  pages={204--213},
  year={2021},
  publisher={SAGE Publications Sage CA: Los Angeles, CA}
}

@article{ma2020bike,
  title={Bike-sharing systems’ impact on modal shift: A case study in Delft, the Netherlands},
  author={Ma, Xinwei and Yuan, Yufei and Van Oort, Niels and Hoogendoorn, Serge},
  journal={Journal of cleaner production},
  volume={259},
  pages={120846},
  year={2020},
  publisher={Elsevier}
}

@article{grau2024cycling,
  title={Cycling into the workshop: e-bike and m-bike mobility patterns for predictive maintenance in Barcelona’s bike-sharing system},
  author={Grau-Escolano, Jordi and Bassolas, Aleix and Vicens, Julian},
  journal={EPJ Data Science},
  volume={13},
  number={1},
  pages={1--21},
  year={2024},
  publisher={SpringerOpen}
}

@article{fan2020dockless,
  title={Dockless bike sharing alleviates road congestion by complementing subway travel: Evidence from Beijing},
  author={Fan, Yichun and Zheng, Siqi},
  journal={Cities},
  volume={107},
  pages={102895},
  year={2020},
  publisher={Elsevier}
}

@article{otero2018health,
  title={Health impacts of bike sharing systems in Europe},
  author={Otero, Isabel and Nieuwenhuijsen, Mark J and Rojas-Rueda, David},
  journal={Environment international},
  volume={115},
  pages={387--394},
  year={2018},
  publisher={Elsevier}
}

@article{clockston2021health,
  title={Health impacts of bike-sharing systems in the US},
  author={Clockston, Raeven Lynn M and Rojas-Rueda, David},
  journal={Environmental research},
  volume={202},
  pages={111709},
  year={2021},
  publisher={Elsevier}
}

@article{ricci2015bike,
  title={Bike sharing: A review of evidence on impacts and processes of implementation and operation},
  author={Ricci, Miriam},
  journal={Research in Transportation Business \& Management},
  volume={15},
  pages={28--38},
  year={2015},
  publisher={Elsevier}
}

@article{zaltz2013structure,
  title={The structure of spatial networks and communities in bicycle sharing systems},
  author={Zaltz Austwick, Martin and O’Brien, Oliver and Strano, Emanuele and Viana, Matheus},
  journal={PloS one},
  volume={8},
  number={9},
  pages={e74685},
  year={2013},
  publisher={Public Library of Science San Francisco, USA}
}

@article{kou2019understanding,
  title={Understanding bike sharing travel patterns: An analysis of trip data from eight cities},
  author={Kou, Zhaoyu and Cai, Hua},
  journal={Physica A: Statistical Mechanics and its Applications},
  volume={515},
  pages={785--797},
  year={2019},
  publisher={Elsevier}
}

@article{chen2017understanding,
  title={Understanding bike trip patterns leveraging bike sharing system open data},
  author={Chen, Longbiao and Ma, Xiaojuan and Nguyen, Thi-Mai-Trang and Pan, Gang and Jakubowicz, J{\'e}r{\'e}mie},
  journal={Frontiers of computer science},
  volume={11},
  pages={38--48},
  year={2017},
  publisher={Springer}
}

@article{li2020understanding,
  title={Understanding intra-urban human mobility through an exploratory spatiotemporal analysis of bike-sharing trajectories},
  author={Li, Wenwen and Wang, Shaohua and Zhang, Xiaoyi and Jia, Qingren and Tian, Yuanyuan},
  journal={International Journal of Geographical Information Science},
  volume={34},
  number={12},
  pages={2451--2474},
  year={2020},
  publisher={Taylor \& Francis}
}

@article{tran2015modeling,
  title={Modeling bike sharing system using built environment factors},
  author={Tran, Tien Dung and Ovtracht, Nicolas and d’Arcier, Bruno Faivre},
  journal={Procedia Cirp},
  volume={30},
  pages={293--298},
  year={2015},
  publisher={Elsevier}
}

@article{elAssi2017effects,
  title={Effects of built environment and weather on bike sharing demand: a station level analysis of commercial bike sharing in Toronto},
  author={Wafic El-Assi and Mohamed Salah Mahmoud and Khandker Nurul Habib},
  journal={Transportation},
  year={2017},
  volume={44},
  pages={589-613},
  url={https://api.semanticscholar.org/CorpusID:133307286}
}

@inproceedings{yang2016mobility,
  title={Mobility modeling and prediction in bike-sharing systems},
  author={Yang, Zidong and Hu, Ji and Shu, Yuanchao and Cheng, Peng and Chen, Jiming and Moscibroda, Thomas},
  booktitle={Proceedings of the 14th annual international conference on mobile systems, applications, and services},
  pages={165--178},
  year={2016}
}

@article{Kim2018investigation,
  title={Investigation on the effects of weather and calendar events on bike-sharing according to the trip patterns of bike rentals of stations},
  author={Kyoungok Kim},
  journal={Journal of Transport Geography},
  year={2018},
  volume={66},
  pages={309-320},
  url={https://api.semanticscholar.org/CorpusID:158630089}
}

@inproceedings{kim2020anatomy,
  title={The anatomy of the daily usage of bike sharing systems: elevation, distance and seasonality},
  author={Kim, Injung and Pelechrinis, Konstantinos},
  booktitle={ACM SIGKDD workshop on Urban Computing},
  year={2020}
}

@article{ashqar2019modeling,
  title={Modeling bike counts in a bike-sharing system considering the effect of weather conditions},
  author={Ashqar, Huthaifa I and Elhenawy, Mohammed and Rakha, Hesham A},
  journal={Case studies on transport policy},
  volume={7},
  number={2},
  pages={261--268},
  year={2019},
  publisher={Elsevier}
}

@inproceedings{singhvi2015predicting,
  title={Predicting bike usage for new york city’s bike sharing system},
  author={Singhvi, Divya and Singhvi, Somya and Frazier, Peter I and Henderson, Shane G and O'Mahony, Eoin and Shmoys, David B and Woodard, Dawn B},
  booktitle={Workshops at the twenty-ninth AAAI conference on artificial intelligence},
  year={2015}
}

@inproceedings{li2015traffic,
  title={Traffic prediction in a bike-sharing system},
  author={Li, Yexin and Zheng, Yu and Zhang, Huichu and Chen, Lei},
  booktitle={Proceedings of the 23rd SIGSPATIAL international conference on advances in geographic information systems},
  pages={1--10},
  year={2015}
}

@inproceedings{chen2016dynamic,
  title={Dynamic cluster-based over-demand prediction in bike sharing systems},
  author={Chen, Longbiao and Zhang, Daqing and Wang, Leye and Yang, Dingqi and Ma, Xiaojuan and Li, Shijian and Wu, Zhaohui and Pan, Gang and Nguyen, Thi-Mai-Trang and Jakubowicz, J{\'e}r{\'e}mie},
  booktitle={Proceedings of the 2016 ACM International Joint Conference on Pervasive and Ubiquitous Computing},
  pages={841--852},
  year={2016}
}

@article{albuquerque2021machine,
  title={Machine learning approaches to bike-sharing systems: A systematic literature review},
  author={Albuquerque, Vit{\'o}ria and Sales Dias, Miguel and Bacao, Fernando},
  journal={ISPRS International Journal of Geo-Information},
  volume={10},
  number={2},
  pages={62},
  year={2021},
  publisher={MDPI}
}

@article{yang2020using,
  title={Using graph structural information about flows to enhance short-term demand prediction in bike-sharing systems},
  author={Yang, Yuanxuan and Heppenstall, Alison and Turner, Andy and Comber, Alexis},
  journal={Computers, Environment and Urban Systems},
  volume={83},
  pages={101521},
  year={2020},
  publisher={Elsevier}
}

@article{almannaa2020dynamic,
  title={Dynamic linear models to predict bike availability in a bike sharing system},
  author={Almannaa, Mohammed H and Elhenawy, Mohammed and Rakha, Hesham A},
  journal={International journal of sustainable transportation},
  volume={14},
  number={3},
  pages={232--242},
  year={2020},
  publisher={Taylor \& Francis}
}

@inproceedings{singla2015incentivizing,
  title={Incentivizing users for balancing bike sharing systems},
  author={Singla, Adish and Santoni, Marco and Bart{\'o}k, G{\'a}bor and Mukerji, Pratik and Meenen, Moritz and Krause, Andreas},
  booktitle={Proceedings of the AAAI Conference on Artificial Intelligence},
  volume={29},
  number={1},
  year={2015}
}

@article{faghih2017empirical,
  title={An empirical analysis of bike sharing usage and rebalancing: Evidence from Barcelona and Seville},
  author={Faghih-Imani, Ahmadreza and Hampshire, Robert and Marla, Lavanya and Eluru, Naveen},
  journal={Transportation Research Part A: Policy and Practice},
  volume={97},
  pages={177--191},
  year={2017},
  publisher={Elsevier}
}

@article{raviv2013static,
  title={Static repositioning in a bike-sharing system: models and solution approaches},
  author={Raviv, Tal and Tzur, Michal and Forma, Iris A},
  journal={EURO Journal on Transportation and Logistics},
  volume={2},
  number={3},
  pages={187--229},
  year={2013},
  publisher={Elsevier}
}

@article{dell2014bike,
  title={The bike sharing rebalancing problem: Mathematical formulations and benchmark instances},
  author={Dell'Amico, Mauro and Hadjicostantinou, Eleni and Iori, Manuel and Novellani, Stefano},
  journal={Omega},
  volume={45},
  pages={7--19},
  year={2014},
  publisher={Elsevier}
}

@article{dell2018bike,
  title={The bike sharing rebalancing problem with stochastic demands},
  author={Dell’Amico, Mauro and Iori, Manuel and Novellani, Stefano and Subramanian, Anand},
  journal={Transportation research part B: methodological},
  volume={118},
  pages={362--380},
  year={2018},
  publisher={Elsevier}
}

@article{chiariotti2018dynamic,
  title={A dynamic approach to rebalancing bike-sharing systems},
  author={Chiariotti, Federico and Pielli, Chiara and Zanella, Andrea and Zorzi, Michele},
  journal={Sensors},
  volume={18},
  number={2},
  pages={512},
  year={2018},
  publisher={MDPI}
}

@article{galatoulas2020spatio,
  title={Spatio-temporal trends of e-bike sharing system deployment: A review in Europe, North America and Asia},
  author={Galatoulas, Nikolaos-Fivos and Genikomsakis, Konstantinos N and Ioakimidis, Christos S},
  journal={Sustainability},
  volume={12},
  number={11},
  pages={4611},
  year={2020},
  publisher={MDPI}
}

@article{bielinski2020lessons,
  title={Lessons from implementing a metropolitan electric bike sharing system},
  author={Bieli{\'n}ski, Tomasz and Dopiera{\l}a, {\L}ukasz and Tarkowski, Maciej and Wa{\.z}na, Agnieszka},
  journal={Energies},
  volume={13},
  number={23},
  pages={6240},
  year={2020},
  publisher={MDPI}
}

@article{simsekoglu2019role,
  title={The role of psychological and socio-demographical factors for electric bike use in Norway},
  author={Simsekoglu, {\"O}zlem and Kl{\"o}ckner, Christian A},
  journal={International journal of sustainable transportation},
  volume={13},
  number={5},
  pages={315--323},
  year={2019},
  publisher={Taylor \& Francis}
}

@article{weinert2007transition,
  title={The transition to electric bikes in China: history and key reasons for rapid growth},
  author={Weinert, Jonathan and Ma, Chaktan and Cherry, Christopher},
  journal={Transportation},
  volume={34},
  pages={301--318},
  year={2007},
  publisher={Springer}
}

@article{wamburu2021ride,
  title={Ride substitution using electric bike sharing: feasibility, cost, and carbon analysis},
  author={Wamburu, John and Lee, Stephen and Hajiesmaili, Mohammad H and Irwin, David and Shenoy, Prashant},
  journal={Proceedings of the ACM on Interactive, Mobile, Wearable and Ubiquitous Technologies},
  volume={5},
  number={1},
  pages={1--28},
  year={2021},
  publisher={ACM New York, NY, USA}
}

@article{campbell2016factors,
  title={Factors influencing the choice of shared bicycles and shared electric bikes in Beijing},
  author={Campbell, Andrew A and Cherry, Christopher R and Ryerson, Megan S and Yang, Xinmiao},
  journal={Transportation research part C: emerging technologies},
  volume={67},
  pages={399--414},
  year={2016},
  publisher={Elsevier}
}

@article{siman2018look,
  title={A look at electric bike casualties: Do they differ from the mechanical bicycle?},
  author={Siman-Tov, Maya and Radomislensky, Irina and Peleg, Kobi and Bahouth, H and Becker, A and Jeroukhimov, I and Karawani, I and Kessel, B and Klein, Y and Lin, G and others},
  journal={Journal of Transport \& Health},
  volume={11},
  pages={176--182},
  year={2018},
  publisher={Elsevier}
}

@article{ji2014electric,
  title={Electric bike sharing: simulation of user demand and system availability},
  author={Ji, Shuguang and Cherry, Christopher R and Han, Lee D and Jordan, David A},
  journal={Journal of Cleaner Production},
  volume={85},
  pages={250--257},
  year={2014},
  publisher={Elsevier}
}

@article{florez2018development,
  title={Development of a bike-sharing system based on pedal-assisted electric bicycles for bogota city},
  author={Florez, David and Carrillo, Henry and Gonzalez, Ricardo and Herrera, Max and Hurtado-Velasco, Ronald and Cano, Martha and Roa, Sergio and Manrique, Tatiana},
  journal={Electronics},
  volume={7},
  number={11},
  pages={337},
  year={2018},
  publisher={MDPI}
}

@article{zhang2019battery,
  title={Battery maintenance of pedelec sharing system: Big data based usage prediction and replenishment scheduling},
  author={Zhang, Chaofeng and Dong, Mianxiong and Luan, Tom H and Ota, Kaoru},
  journal={IEEE Transactions on Network Science and Engineering},
  volume={7},
  number={1},
  pages={127--138},
  year={2019},
  publisher={IEEE}
}

@article{usama2019towards,
  title={Towards an energy efficient solution for bike-sharing rebalancing problems: a battery electric vehicle scenario},
  author={Usama, Muhammad and Shen, Yongjun and Zahoor, Onaira},
  journal={Energies},
  volume={12},
  number={13},
  pages={2503},
  year={2019},
  publisher={MDPI}
}

@article{zhou2023dynamic,
  title={Dynamic battery swapping and rebalancing strategies for e-bike sharing systems},
  author={Zhou, Yaoming and Lin, Zeyu and Guan, Rui and Sheu, Jiuh-Biing},
  journal={Transportation Research Part B: Methodological},
  volume={177},
  pages={102820},
  year={2023},
  publisher={Elsevier}
}

@article{soriguera2018simulation,
  title={A simulation model for public bike-sharing systems},
  author={Soriguera, Francesc and Casado, V{\'\i}ctor and Jim{\'e}nez, Enrique},
  journal={Transportation research procedia},
  volume={33},
  pages={139--146},
  year={2018},
  publisher={Elsevier}
}

@article{winslow2019bicycle,
  title={Bicycle sharing: Sustainable value creation and institutionalisation strategies in Barcelona},
  author={Winslow, Julia and Mont, Oksana},
  journal={Sustainability},
  volume={11},
  number={3},
  pages={728},
  year={2019},
  publisher={MDPI}
}

@article{burani2022algorithm,
  title={An algorithm to predict e-bike power consumption based on planned routes},
  author={Burani, Erik and Cabri, Giacomo and Leoncini, Mauro},
  journal={Electronics},
  volume={11},
  number={7},
  pages={1105},
  year={2022},
  publisher={MDPI}
}

@article{steyn2014comparison,
  title={Comparison of tyre rolling resistance for different mountain bike tyre diameters and surface conditions},
  author={Steyn, Wynand J vdM and Warnich, Janike},
  journal={South African Journal for Research in Sport, Physical Education and Recreation},
  volume={36},
  number={2},
  pages={179--193},
  year={2014},
  publisher={North-West University}
}

@article{lu2013approaching,
  title={Approaching the limit of predictability in human mobility},
  author={Lu, Xin and Wetter, Erik and Bharti, Nita and Tatem, Andrew J and Bengtsson, Linus},
  journal={Scientific reports},
  volume={3},
  number={1},
  pages={2923},
  year={2013},
  publisher={Nature Publishing Group UK London}
}

@article{huang2015predicting,
  title={Predicting human mobility with activity changes},
  author={Huang, Wei and Li, Songnian and Liu, Xintao and Ban, Yifang},
  journal={International Journal of Geographical Information Science},
  volume={29},
  number={9},
  pages={1569--1587},
  year={2015},
  publisher={Taylor \& Francis}
}

@misc{OpenStreetMap,
    author = "{OpenStreetMap contributors}",
    title = {Planet dump retrieved from https://planet.osm.org},
    howpublished = {\url{https://www.openstreetmap.org/}},
    note = {Accessed: 2024-08-20} ,
    year=2013,
}

@misc{opentopodata,
    author = {Nisbet, A.},
    title = {OpenTopoData},
    howpublished = {\url{https://github.com/ajnisbet/opentopodata}},
    note = {Accessed: 2024-08-20},
    year = {2013},
}

@misc{openrouteservice,
    author = "{GIScience Research Group and HeiGIT}",
    title = {openrouteservice},
    howpublished = {\url{https://openrouteservice.org/}},
    note = {Accessed: 2024-08-20} ,
    year=2023,
}

@inproceedings{akiba2019optuna,
  title={Optuna: A next-generation hyperparameter optimization framework},
  author={Akiba, Takuya and Sano, Shotaro and Yanase, Toshihiko and Ohta, Takeru and Koyama, Masanori},
  booktitle={Proceedings of the 25th ACM SIGKDD international conference on knowledge discovery \& data mining},
  pages={2623--2631},
  year={2019}
}

@article{bertucci2013evaluation,
  title={Evaluation of aerodynamic and rolling resistances in mountain-bike field conditions},
  author={Bertucci, William M and Rogier, Simon and Reiser, Raoul F},
  journal={Journal of sports sciences},
  volume={31},
  number={14},
  pages={1606--1613},
  year={2013},
  publisher={Taylor \& Francis}
}

@misc{bcn_population_density,
    author = {{Ajuntament de Barcelona}},
    title = {{Population density of Barcelona neighborhoods}},
    howpublished = {\url{https://ajuntament.barcelona.cat/estadistica/angles/Estadistiques_per_temes/Medi_urba/Territori/Superficie/a2021/S0403.htm}},
    note = {Accessed: 2024-09-09},
    year = {2021},
}

\clearpage

\setcounter{figure}{0}
\setcounter{table}{0}
\setcounter{section}{0}

\renewcommand{\thefigure}{S\arabic{figure}}
\renewcommand{\thetable}{S\arabic{table}}
\renewcommand{\thesubsection}{S\arabic{section}}  
\renewcommand{\theequation}{S\arabic{equation}} 
\renewcommand{\theHfigure}{S\arabic{figure}}
\renewcommand{\theHtable}{S\arabic{table}}

\setcounter{equation}{0}

\section*{Supplementary Material \\ Spatiotemporal variability and prediction of e-bike battery levels in bike-sharing systems}

\section{Data description}

\begin{figure}[!htbp]
  \begin{center}
  \includegraphics[width=0.95\textwidth]{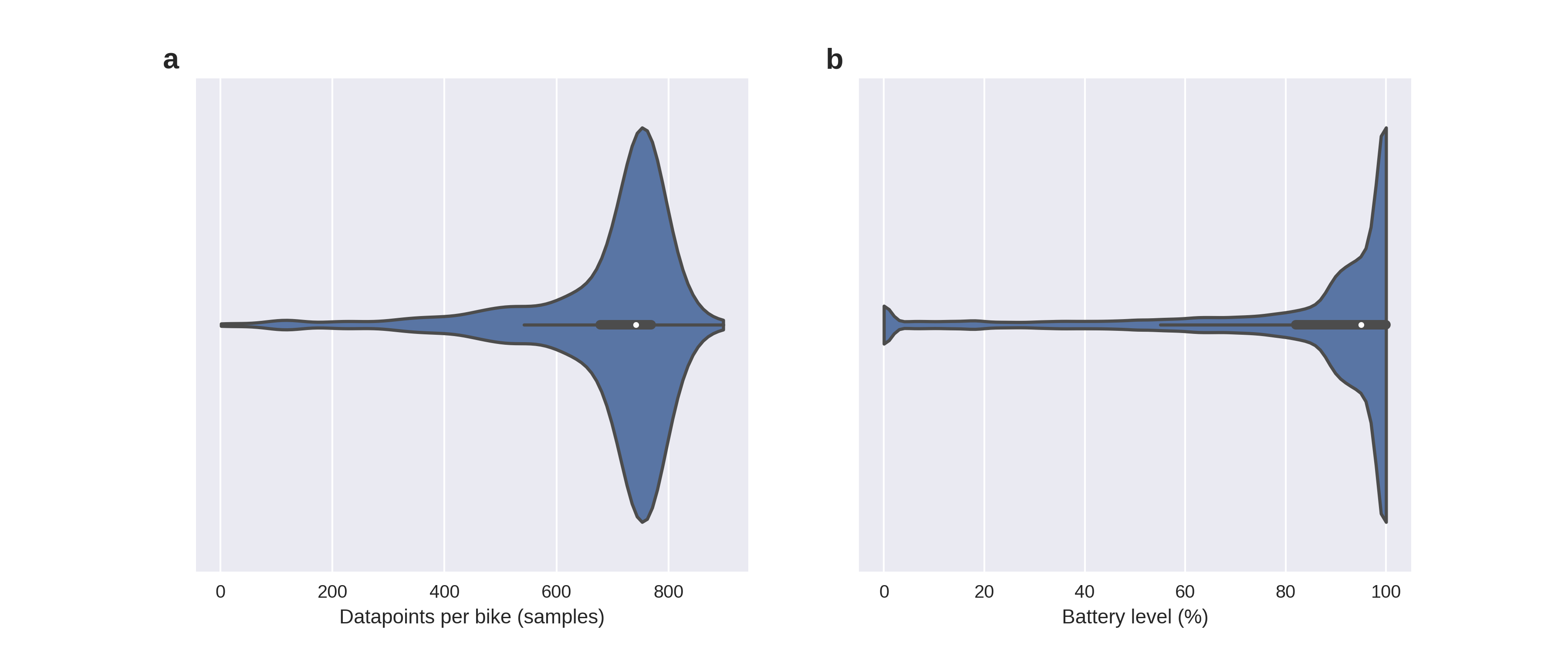}
  \end{center}
  \caption[30 minutes dataset description]{\textbf{30-minute dataset description.} \textbf{a.} Distribution of data points per bike. \textbf{b.} Distribution of battery levels across the dataset.} \label{fig:30min}
\end{figure}

The first dataset we analysed comprises battery levels recorded every 8 hours, containing data on 2,748 bikes from July 20th to December 31st, 2022 (spanning 164 days). Battery levels are recorded exclusively when bikes are docked at stations, resulting in 1.48 million data points. The dataset fields are datetime in second granularity, bike identifier, station, status (operative or inoperative for various maintenance reasons), and battery metrics such as voltage, temperature, current, and battery level (from 0 to 1). After excluding records with missing values and filtering for those labelled as ‘operative’, the dataset was narrowed to 671,251 rows. Additionally, for the modelling phase, only data points with a subsequent data point for the same bike 8 hours later were utilised. The details are provided in Figure \ref{fig:30min}.

\begin{figure}[!htbp]
  \begin{center}
  \includegraphics[width=0.95\textwidth]{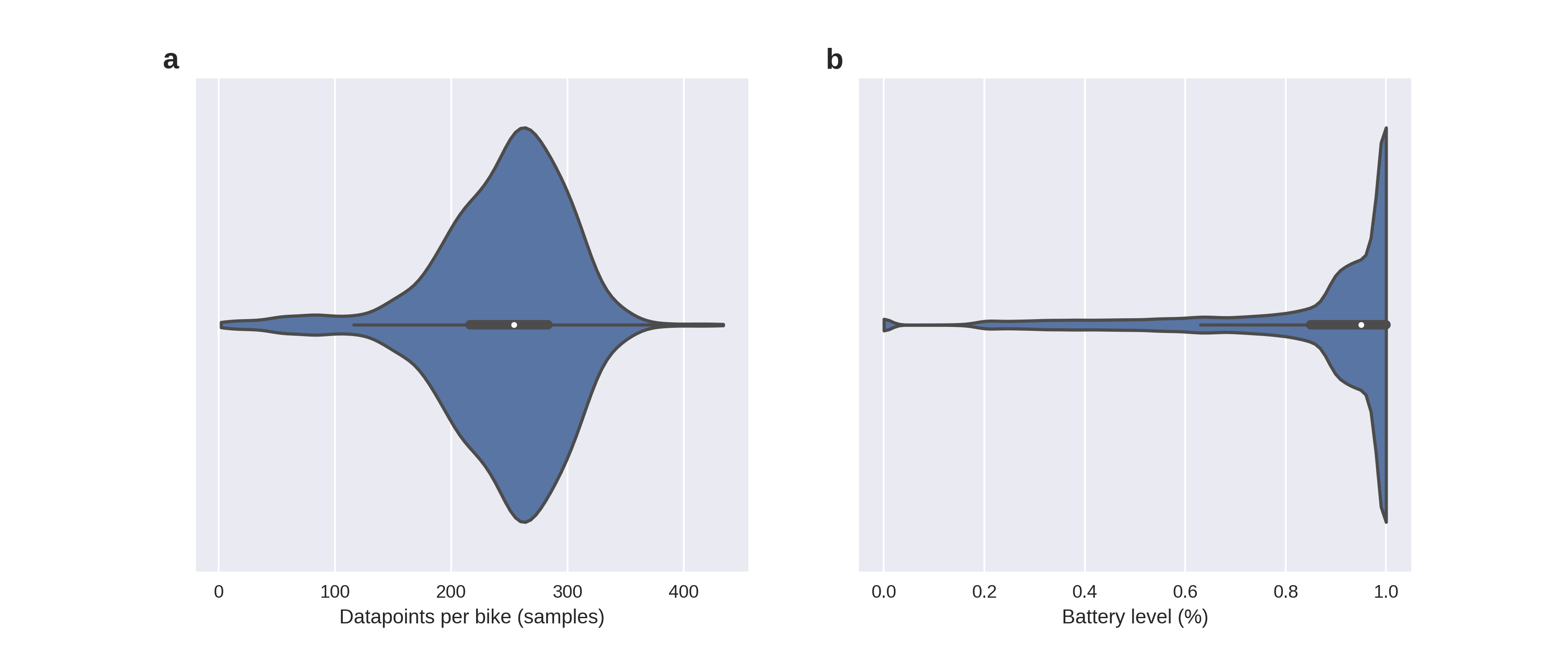}
  \end{center}
  \caption[8 hours dataset data description]{\textbf{8-hour dataset data description.} \textbf{a.} Distribution of data points per bike. \textbf{b.} Distribution of battery levels across the dataset.} \label{fig:8hour}
\end{figure}

The second dataset contains battery level snapshots taken every 30 minutes for 3,004 bikes, totalling 2.71 million data points from March 29th to April 18th, 2023 (spanning 20 days). Similarly, this dataset includes the same fields, with battery level information available only when bikes are docked. After data cleaning and selecting the bikes with an ‘operative’ status, 1.87 million data points remained. For analysis, only data points with a subsequent reading for the same bike 8 hours later were used. The details are provided in Figure \ref{fig:8hour}. 

\section{Data analysis}

\subsection*{Battery levels across time for each neighbourhood}

\begin{figure}[H]
    \centering
    \includegraphics[width=0.95\textwidth]{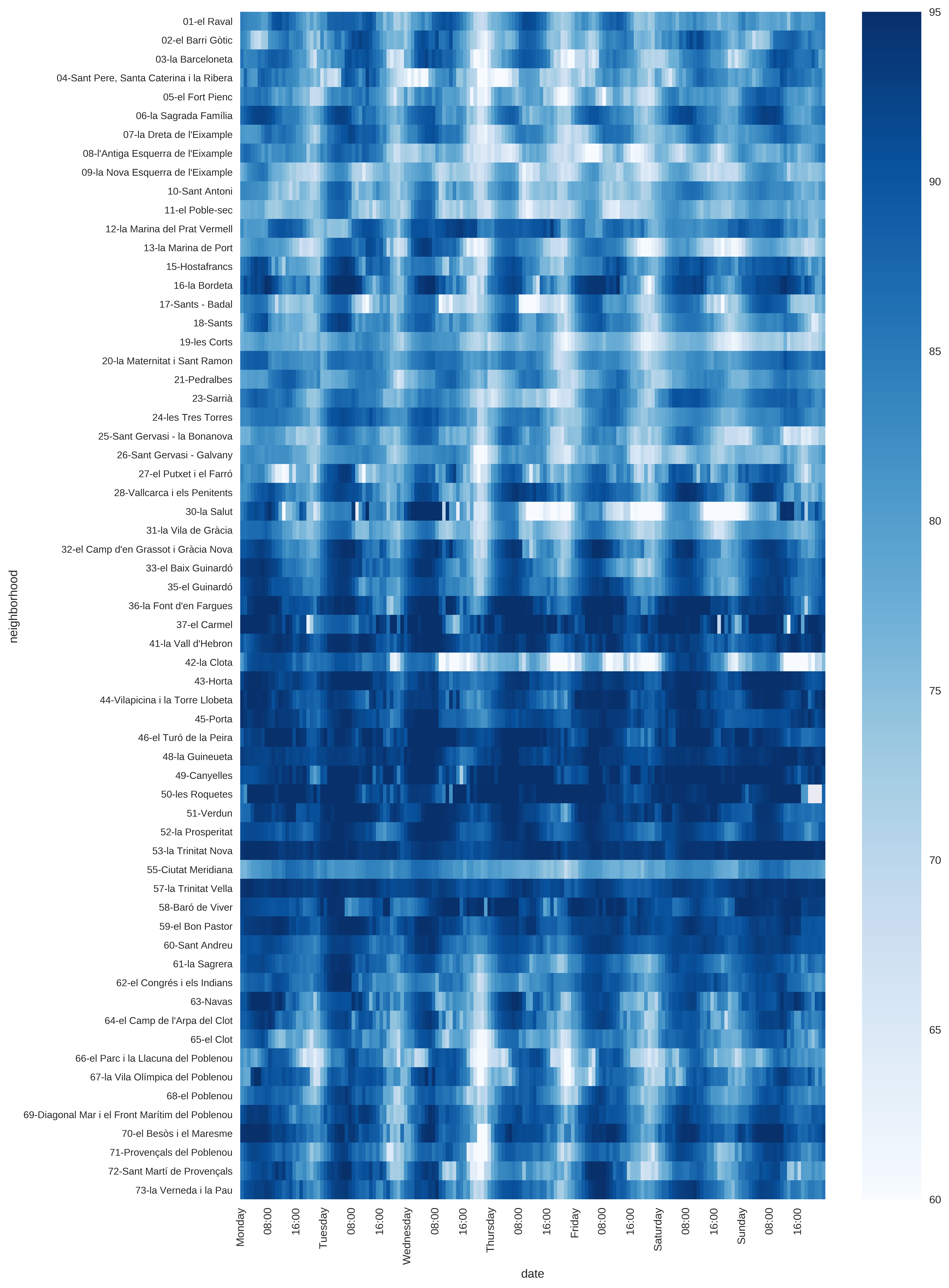}
    \caption[Average battery levels across time for each neighbourhood]{\textbf{Average battery levels across time for each neighbourhood.} The x-axis represents time for an average week, while the y-axis lists the neighbourhoods.}
    \label{fig:avg_battery_neighborhood_heatmap}
\end{figure}

\subsection*{Choice of number of clusters}

\begin{figure}[H]
    \centering
    \includegraphics[width=0.9\textwidth]{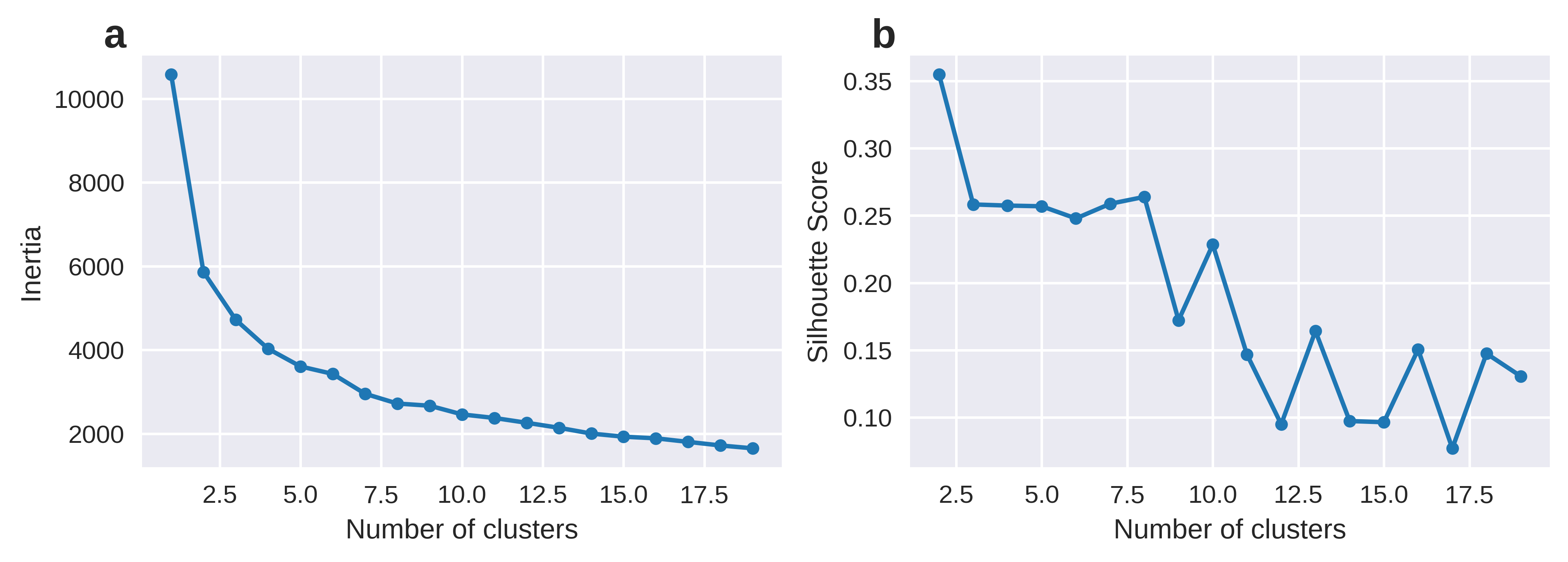}
    \caption[Determining the optimal number of clusters]{\textbf{Determining the optimal number of clusters.} \textbf{a.} Elbow Method for optimal k. \textbf{b.} Silhouette Method  for optimal k.}
    \label{fig:find_k}
\end{figure}

To determine the optimal number of clusters for the k-means clustering analysis, we utilized two commonly used methods: the Elbow Method and the Silhouette Method (Figure~\ref{fig:find_k}). The Elbow Method helps identify the point where the within-cluster sum of squares (inertia) begins to decrease at a slower rate, indicating the optimal number of clusters. The Silhouette Method measures the quality of the clustering by calculating the average silhouette score for different numbers of clusters.

Based on these analyses, we chose to use 4 clusters for our k-means clustering. This choice balances the trade-off between minimising within-cluster variance (as indicated by the Elbow Method) and maximising cluster separation (as indicated by the Silhouette Method).

\subsection*{Collinearity between inter-event time and distance index}

To ensure that there is no significant collinearity between the inter-event time and the distance index, we examined both the correlation matrix and the Variance Inflation Factor (VIF) for each variable in the regression model.

\begin{figure}[H]
    \centering
    \includegraphics[width=0.6\textwidth]{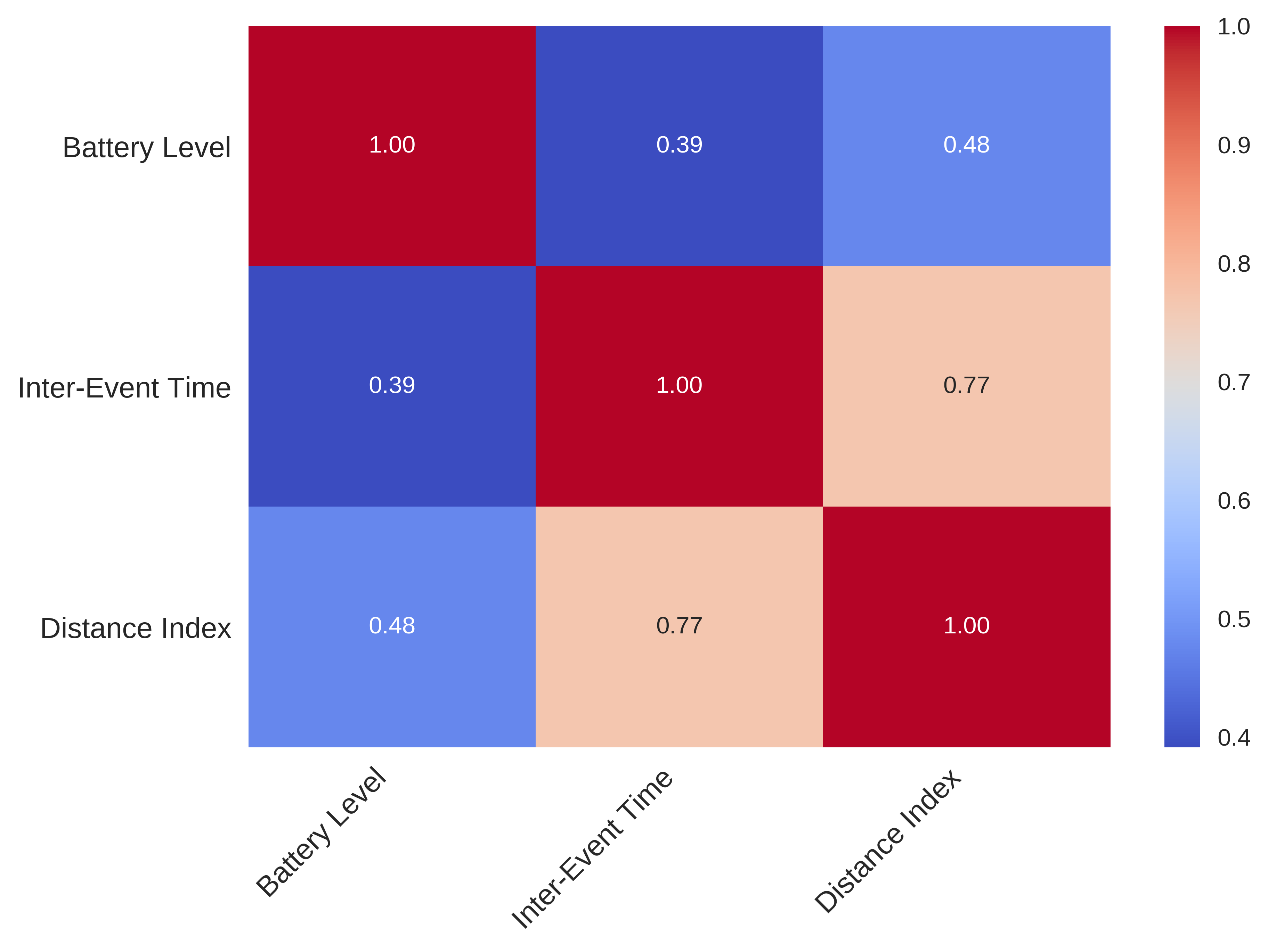}
    \caption[Correlation matrix for battery level, inter-event time, and distance index]{\textbf{Correlation matrix for battery level, inter-event time, and distance index.}}
    \label{fig:correlation_matrix}
\end{figure}

Figure~\ref{fig:correlation_matrix} illustrates the correlation values among average battery level, inter-event time, and distance index. The correlation between inter-event time and distance index is 0.77, indicating a moderate relationship. At first glance, this might suggest potential collinearity between these two variables.

However, to confirm whether this correlation leads to collinearity, we computed the VIF for each variable. The VIF quantifies how much the variance of a regression coefficient is inflated due to collinearity with other predictors. A VIF value greater than 10 typically indicates high collinearity.

The table below summarises the VIF values for the variables:

\begin{table}[H]
    \centering
    \begin{tabular}{lc}
        \toprule
        Variable & VIF \\
        \midrule
        constant & 170.328236 \\
        battery level & 1.296030 \\
        inter-event time & 2.447021 \\
        distance index & 2.680041 \\
        \bottomrule
    \end{tabular}
    \caption{\textbf{Variance Inflation Factor (VIF) values for the variables in the regression model.}}
    \label{tab:vif}
\end{table}

As shown in Table~\ref{tab:vif}, the VIF values for the inter-event time and the distance index are 2.447021 and 2.680041, respectively. These values are well below the threshold of 10, indicating that there is no significant collinearity between these variables.

Thus, considering both the correlation matrix and the VIF values, we conclude that the inter-event time and the distance index can be used as independent predictors in the regression model without concerns of collinearity.

\section{Battery validation and elevation data}

\begin{figure}[!htbp]
  \begin{center}
  \includegraphics[width=0.9\textwidth]{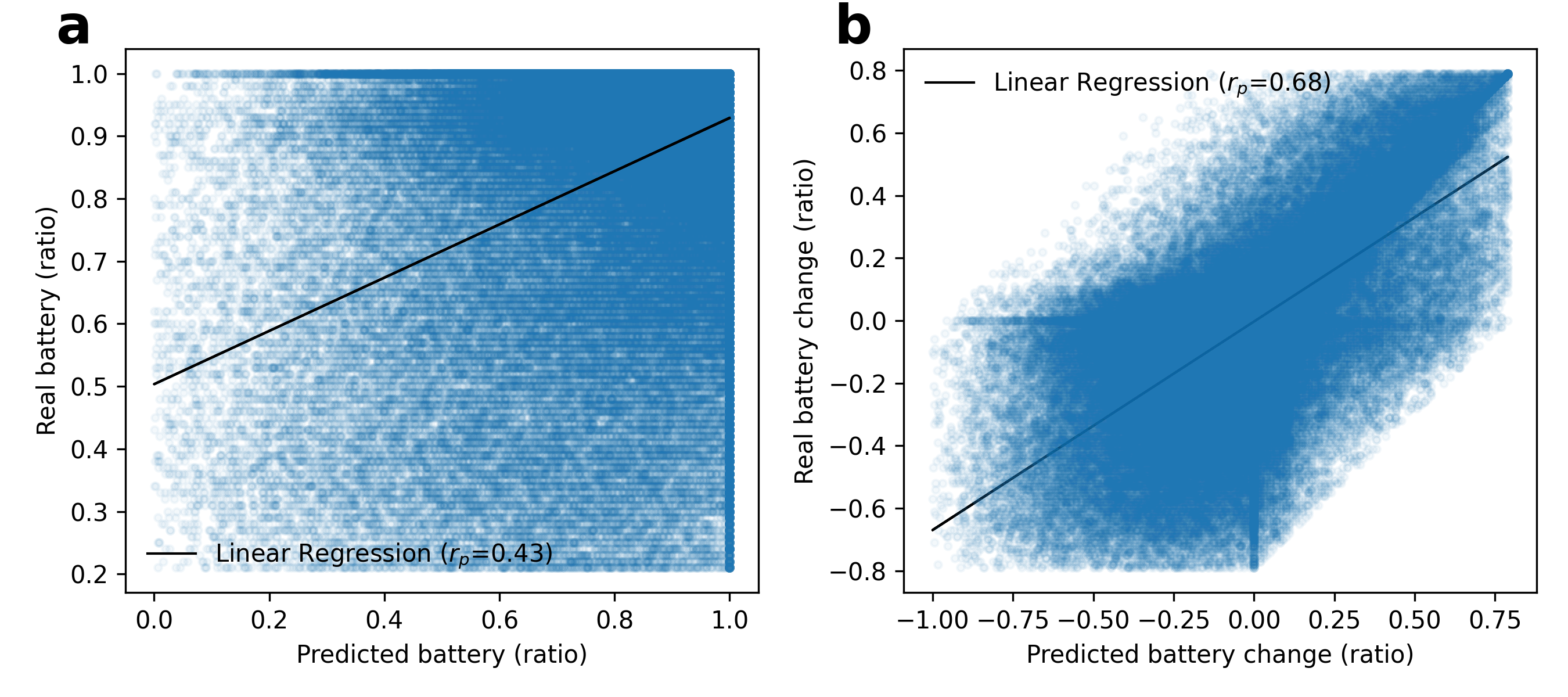}
\end{center}
  \caption[Battery model fit in the 8-hour dataset]{\textbf{Battery model fit in the 8-hour dataset.} \textbf{a.} Observed battery level as a function of the predicted values. \textbf{b.} Battery change observed as a function of the predicted values. The Root Mean Squared Error (RMSE) is $0.036$. The linear fit between both is shown in grey.} \label{fig:8hour_validation}
\end{figure}

In Figure \ref{fig:8hour_validation}, the correlation between the predicted and observed battery levels in the 8-hour dataset is shown.

\begin{figure}[!htbp]
  \begin{center}
  \includegraphics[width=0.9\textwidth]{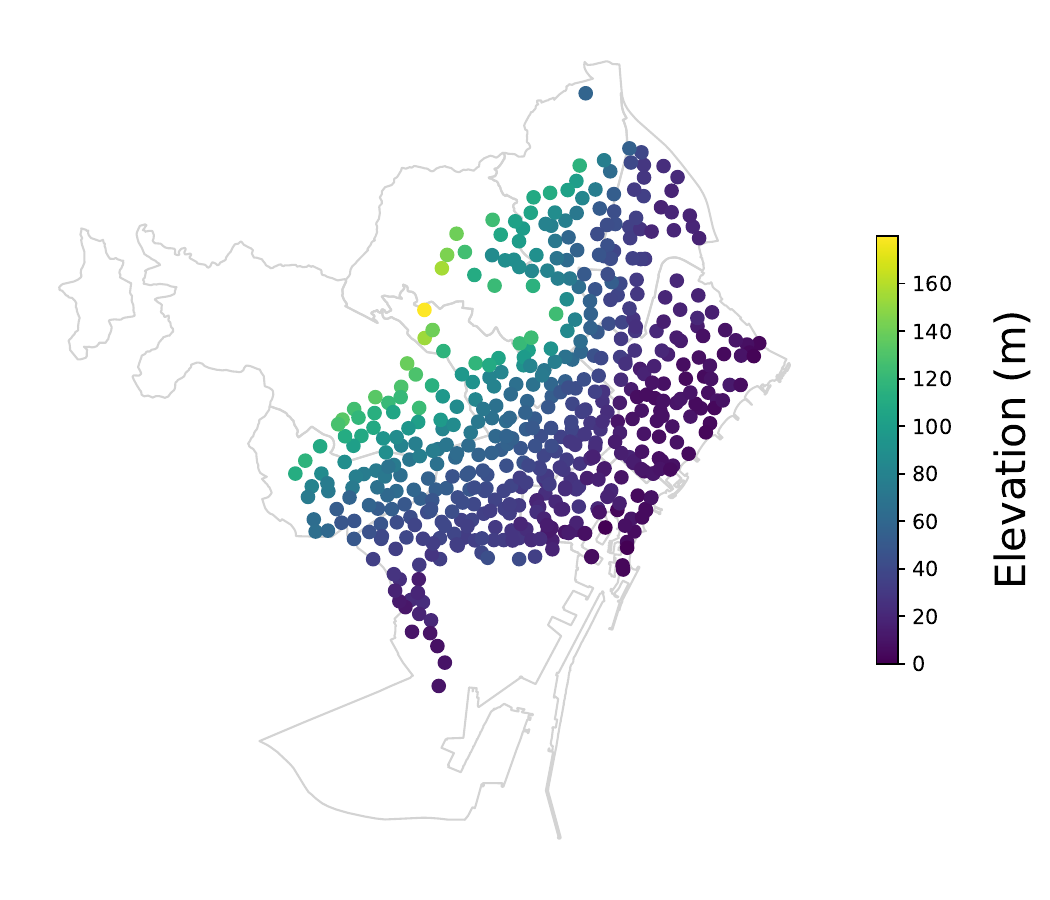}
\end{center}
  \caption[Station altitude]{\textbf{Station altitude.} Altitude of the Bicing stations.} \label{fig:elevation}
\end{figure}

Battery consumption is highly impacted by the altitude difference between the origin and destination stations of a trip. The altitude of each station in the BSS of Barcelona is provided in Figure \ref{fig:elevation}.
A clear division between the station near the sea and those in the northern region is observed. In some cases, the disparity between stations is steep, with differences in altitude greater than 100 metres.

\end{document}